\documentclass[aps, prd, amsmath, floats, floatfix, twocolumn, nofootinbib, showpacs]{revtex4-1}
\usepackage{graphicx}
\usepackage{color}

\newcommand{\be}{\begin{eqnarray}}
\newcommand{\ee}{\end{eqnarray}}

\begin{document}

\title{Testing the Kerr-nature of stellar-mass black hole candidates 
by combining the continuum-fitting method and the power estimate of transient ballistic jets}

\author{Cosimo Bambi}
\email{Cosimo.Bambi@physik.uni-muenchen.de}
\affiliation{Arnold Sommerfeld Center for Theoretical Physics\\
Ludwig-Maximilians-Universit\"at M\"unchen, 80333 Munich, Germany}

\date{\today}

\begin{abstract}
Astrophysical black hole candidates are thought to be the Kerr black holes 
predicted by General Relativity, as these objects cannot be explained otherwise 
without introducing new physics. However, there is no observational evidence 
that the space-time around them is really described by the Kerr solution. The 
Kerr black hole hypothesis can be tested with the already available X-ray data by 
extending the continuum-fitting method, a technique currently used by astronomers 
to estimate the spins of stellar-mass black hole candidates. In general, we cannot 
put a constraint on possible deviations from the Kerr geometry, but only on some 
combination between these deviations and the spin. The measurement of the
radio power of transient jets in black hole binaries can potentially break this
degeneracy, thus allowing for testing the Kerr-nature of these objects.
\end{abstract}

\pacs{97.60.Lf, 97.80.Jp, 04.50.Kd, 97.10.Gz, 98.38.Fs}

\maketitle


\section{Introduction}

The $5-20$~$M_\odot$ compact objects in X-ray binary systems and the
$10^5 - 10^9$~$M_\odot$ dark bodies at the center of every normal 
galaxy are thought to be the Kerr black holes (BHs) predicted by General 
Relativity~\cite{nar}. There is no evidence that the space-time around these 
objects is really described by the Kerr metric, but, at the same time, there 
is no other explanation in the framework of conventional physics. A Kerr BH is 
completely specified by two parameters: its mass, $M$, and its spin angular 
momentum, $J$. A fundamental limit for a BH in 4-dimensional General 
Relativity is the bound $|a_*| \le 1$, where $a_* = J/M^2$ is the dimensionless 
spin parameter\footnote{Throughout the paper, I use units in which 
$G_{\rm N}=c=1$.}. This is just the condition for the existence of the event 
horizon: for $|a_*| > 1$, there is no horizon and the Kerr metric describes a 
naked singularity, which is forbidden by the weak cosmic censorship 
conjecture~\cite{pen}.

In the case of the stellar-mass BH candidates in X-ray binary systems, 
the mass $M$ can be deduced by studying the orbital motion of the stellar 
companion. This measurement is reliable, because the system can be 
described in the framework of Newtonian mechanics, with no assumptions
about the nature of the compact object. The situation changes when we
want to get an estimate of the spin parameter $a_*$. The most reliable approach 
is currently the continuum-fitting method~\cite{cfm0,cfm,bh1,bh2,bh3,bh4}. 
Basically, one fits the X-ray continuum spectrum of the BH candidate using the 
standard accretion disk model of Novikov and Thorne~\cite{nt}. Under the assumption 
that the background geometry is described by the Kerr metric, it is possible 
to infer the spin parameter, $a_*$, and the mass accretion rate, $\dot{M}$, 
if the mass of the BH candidate, its distance from us, and the inclination 
angle of the disk are known independently.

The possibility of testing the Kerr nature of astrophysical BH candidates
with present and near future experiments is becoming an active research
field~\cite{rev,exp,bb,iron,jpp,orbits,gw}. In particular, one can extend the continuum-fitting
method to constrain possible deviations from the Kerr geometry~\cite{bb}.
That can be achieved by considering a more general background, which 
includes the Kerr solution as special case. The compact object will be thus
characterized by $M$, $a_*$, and at least one ``deformation parameter'', 
measuring deviations from the Kerr geometry. If observational data require 
a vanishing deformation parameter, the Kerr BH hypothesis is verified.
However, the fit of the X-ray spectrum cannot be used to measure $a_*$ 
and the deformation parameter at the same time, but it is only possible to 
constrain a combination of them. This is not a problem of the continuum-fitting 
method, but of any approach (see e.g. Ref.~\cite{iron} for the case of the analysis of the 
K$\alpha$ iron line).

In what follows, I will apply the recent finding of Ref.~\cite{nm} to show
that one can potentially break the degeneracy between $a_*$ and the 
deformation parameter by combining the continuum-fitting method with 
the power estimate of transient ballistic jets.

\begin{table*}
\begin{center}
\begin{tabular}{c c c c c c c c c c c}
\hline
\hspace{.5cm} & BH Binary & \hspace{.5cm} & $a_*$ &  \hspace{.5cm} & $\eta$ &  \hspace{.5cm} & $P_{\rm jet}$ (kpc$^2$ GHz Jy/$M_{\odot}$) &  \hspace{.5cm} & Reference &  \hspace{.5cm} \\
\hline 
& GRS 1915+105 & & $0.975$, $a_* > 0.95$ & & $0.224$, $\eta > 0.190$ & & $39.4$ & & \cite{bh1} & \\
& GRO J1655-40 & & $0.7 \pm 0.1$ & & $0.104^{+0.018}_{-0.013}$ & & $19.7$ & & \cite{bh2} & \\ 
& XTE J1550-564 & & $0.34 \pm 0.24$ & & $0.072^{+0.017}_{-0.011}$ & & $2.79$ & & \cite{bh3} & \\
& A0620-00 & & $0.12 \pm 0.19$ & & $0.061^{+0.009}_{-0.007}$ & & $0.173$ & & \cite{bh4} & \\
\hline
\end{tabular}
\end{center}
\vspace{-0.2cm}
\caption{The four stellar-mass BH candidates of which the spin parameter
$a_*$ has been estimated with the continuum-fitting method and we have
radio data of their outbursts. The accretion efficiency $\eta$ in the third 
column has been deduced from the corresponding $a_*$ for a Kerr
background. The mass-normalized jet power $P_{\rm jet}$ in the fourth 
column has been inferred from the data reported in Ref.~\cite{nm},
using Eq.~(\ref{e-j}).}
\label{tab}
\end{table*}

\section{Transient ballistic jets}

Observationally, BH binaries can emit two kinds of jets~\cite{fbg}. {\it Steady jets} 
occur in the hard spectral state, over a wide range of luminosity of the source, 
and they seem to be not very relativistic. {\it Transient ballistic jets} are instead 
launched when a BH binary with a low-mass companion undergoes a transient 
outburst: the jet appears when the source switches from the hard to soft state 
and its luminosity is close to the Eddington limit. Transient jets are observed 
as blobs of plasma moving ballistically outward at relativistic velocities.
The common interpretation is that steady jets are produced relatively far from 
the compact object, say at about 10 to 100 gravitational radii~\cite{j1}, while
transient jets are launched within a few gravitational radii~\cite{j2}. As discussed 
in Ref.~\cite{nm}, it is therefore plausible that transient jets are powered by  
the rotational energy of the BH and, since they occur at a well defined 
luminosity, they may be used as ``standard candles''.

In Ref.~\cite{nm}, the authors show there is a correlation between
the spin parameter $a_*$, as inferred by the continuum-fitting method,
and the radio power of transient ballistic jets. Moreover, the behavior is 
close to what should be expected if these jets were powered by the 
BH spin via the Blandford-Znajek mechanism~\cite{bz}.

So far, the continuum-fitting method has provided the estimate of the spin
parameter of nine stellar-mass BH candidates~\cite{cfm}. Five of these objects
have a low-mass companion and undergo mass transfer via Roche lobe 
outflow: during their outbursts, they produce ballistic jets. For three of them
(GRS 1915+105, GRO J1655-40, and XTE J1550-564), we have good
radio data during at least one of their outbursts. For A0620-00, the data are
not so good. 4U 1543-47 has never been monitored well at radio wavelength
during  any of its outbursts. For GRS 1915+105, GRO J1655-40, XTE J1550-564,
and A0620-00, the authors of Ref.~\cite{nm} compute the mass-normalized 
jet radio power:
\be\label{e-j}
P_{\rm jet} = \frac{D^2 (\nu S_\nu)_{\rm max, 5GHz}}{M} \, ,
\ee
where $D$ is the distance of the binary system from us and
$(\nu S_\nu)_{\rm max, 5GHz}$ is the estimate of the maximum of
the radio power at 5~GHz (see Tab.~\ref{tab}). Then, they plot the jet power 
$P_{\rm jet}$ against the BH spin parameter $a_*$, as inferred from the
continuum-fitting method, and against the corresponding BH angular
frequency
\be\label{e-o}
\Omega_H = - \frac{g_{t\phi}}{g_{\phi\phi}}\Big|_{r = r_H} =
\frac{a}{r_H^2 + a^2} \, ,
\ee
where $r_H$ is the radius of the BH outer event horizon and $a = a_*M$.
The scaling $P_{\rm jet} \sim a_*^2$ was derived in Ref.~\cite{bz},
under the assumption $|a_*| \ll 1$. $P_{\rm jet} \sim \Omega_H^2$
was instead obtained in Ref.~\cite{tnm} and works even for
spin parameters quite close to 1. The top left panel of Fig.~\ref{f-jet}
shows the plot $P_{\rm jet}$ vs $\Omega_H$, which is basically the 
plot in Fig.~3 of Ref.~\cite{nm}. The blue-dashed line has slope
of 2, as expected from the theoretical scaling. The uncertainty in
$P_{\rm jet}$ is the somehow arbitrary uncertainty of 0.3 in the
log adopted in Ref.~\cite{nm}. Despite there being only four objects,
there is evidence for a correlation between jet power and 
$\Omega_H$, and one finds the behavior expected in the
case of a jet powered by the rotational energy of the BH. For more
details about the systematics, the interpretation of the finding, and
the comparison with previous results, see Ref.~\cite{nm}.
The conclusions of the authors are therefore that: $i)$ they have
provided the first evidence that some jets may be powered by the
BH spin energy, and $ii)$ the observed correlation also provides
an additional confirmation of the continuum-fitting method.

\begin{figure*}
\begin{center}  
\includegraphics[type=pdf,ext=.pdf,read=.pdf,width=8cm]{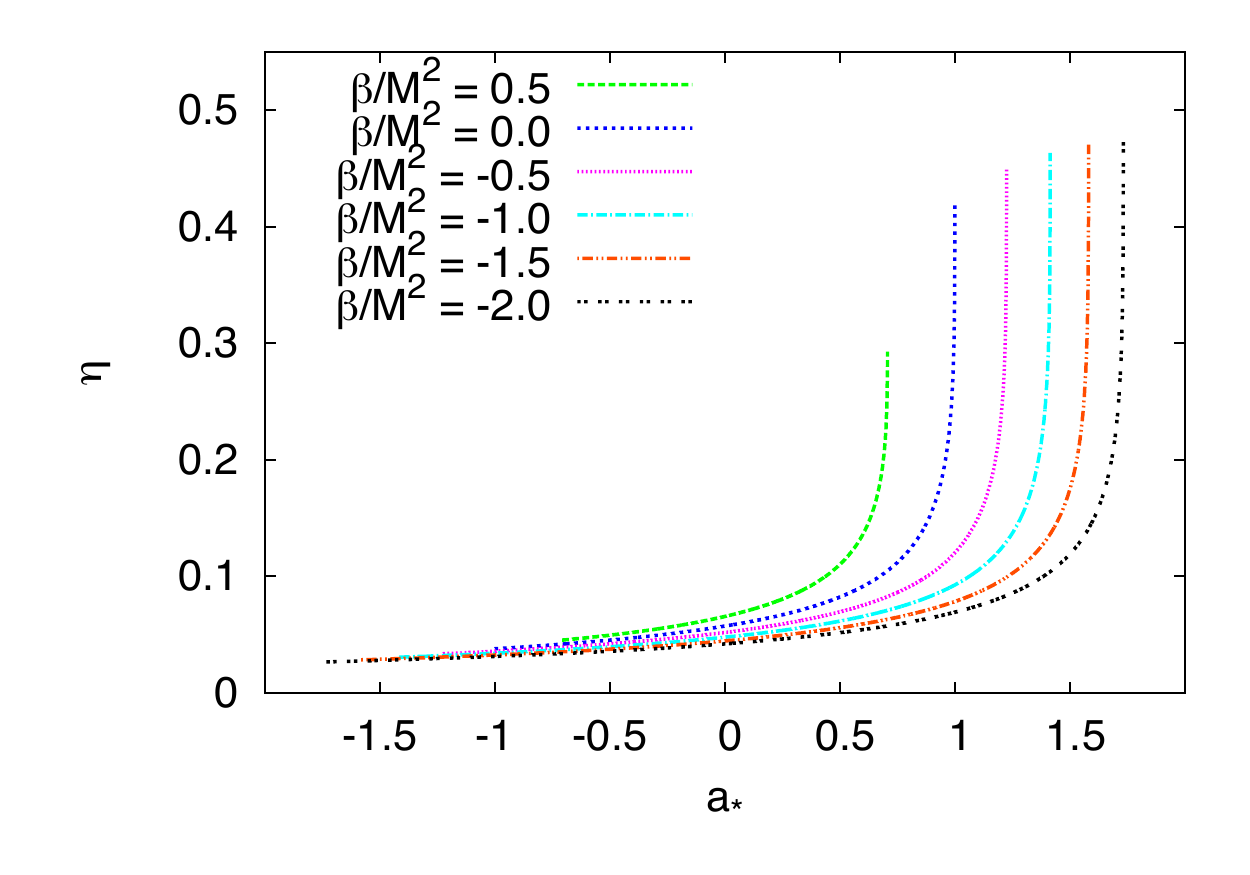}
\includegraphics[type=pdf,ext=.pdf,read=.pdf,width=8cm]{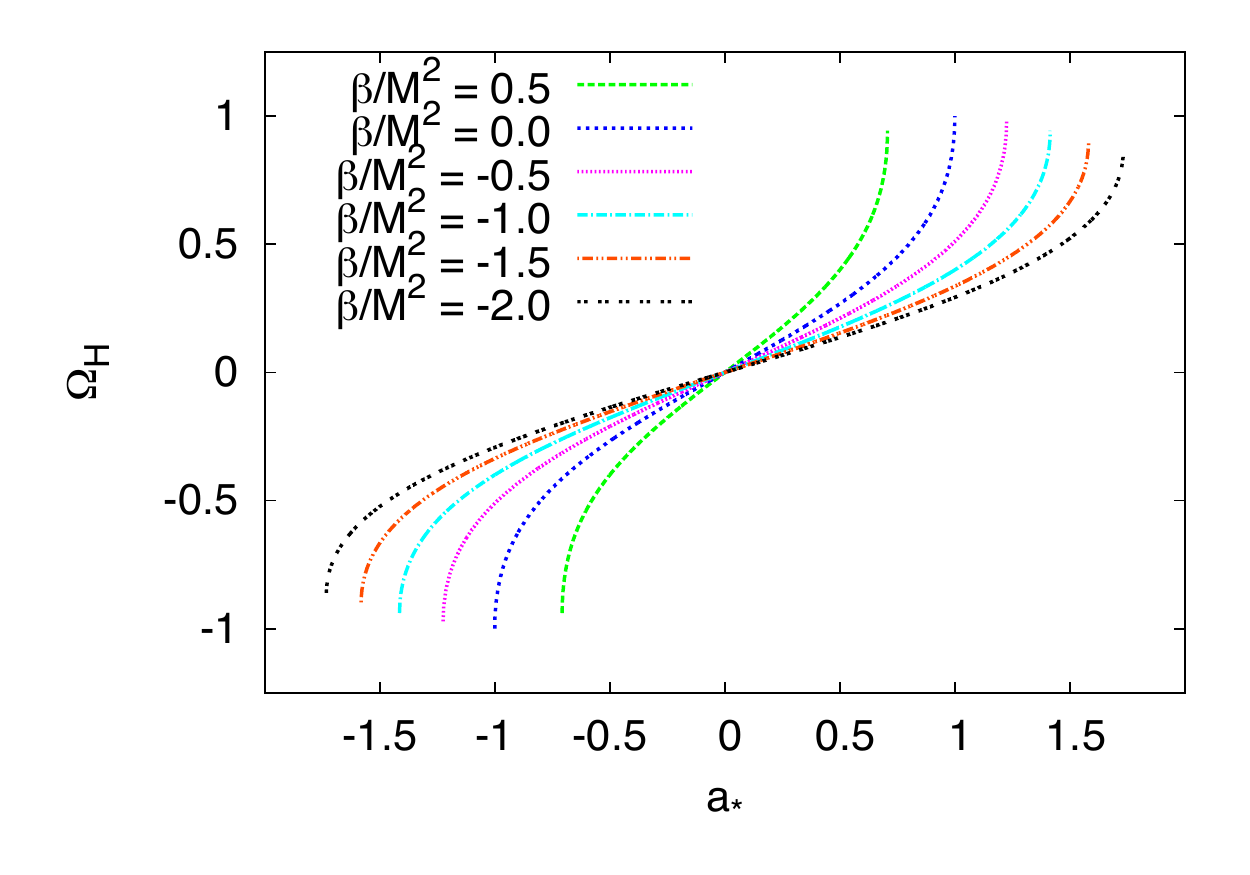}
\vspace{-0.5cm}
\caption{Braneworld-inspired black holes of Eq.~(\ref{e-rs}).
Accretion efficiency $\eta = 1 - E_{\rm ISCO}$ (left panel)
and BH angular frequency $\Omega_H$ (right panel) as a function 
of the spin parameter $a_*$ for different values of $\beta/M^2$.}
\label{f-om}
\includegraphics[type=pdf,ext=.pdf,read=.pdf,width=8cm]{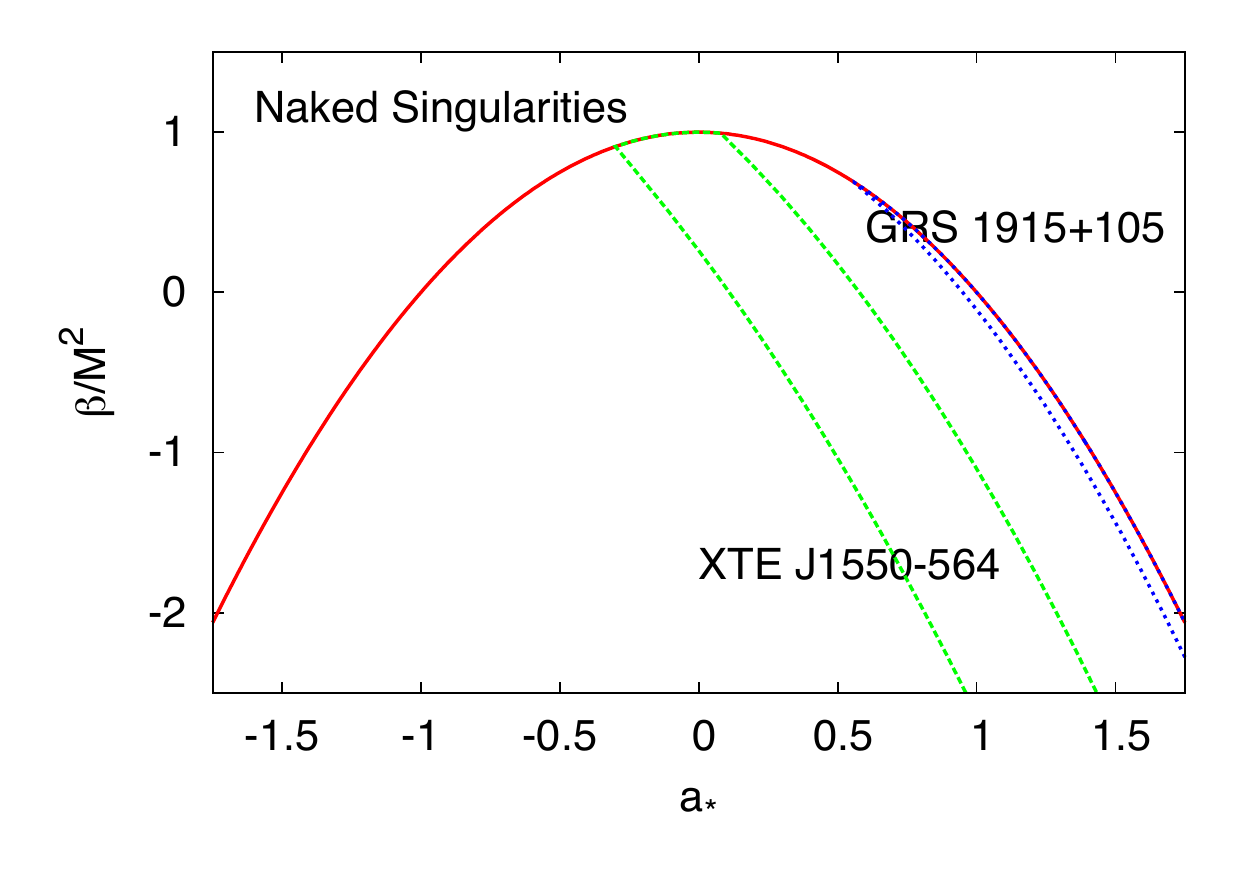}
\includegraphics[type=pdf,ext=.pdf,read=.pdf,width=8cm]{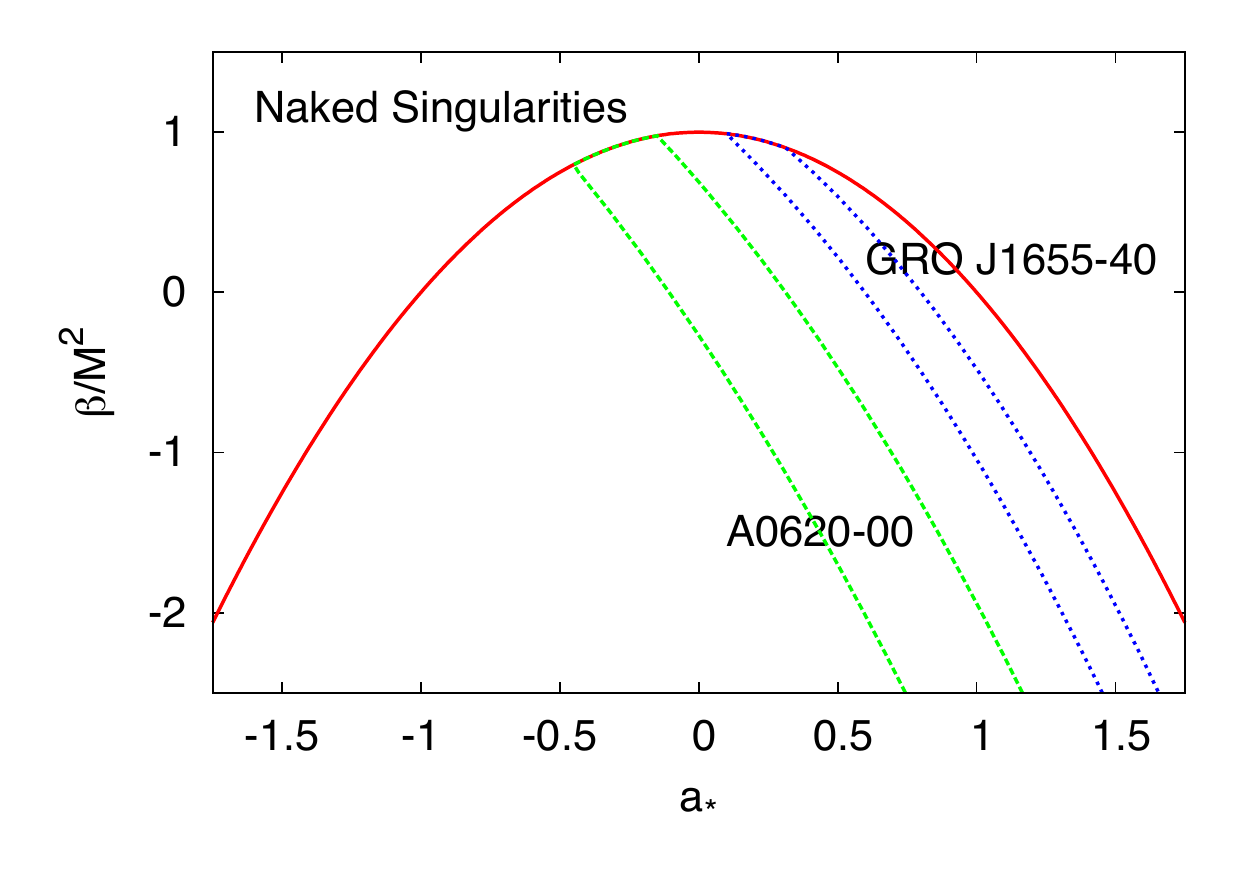}
 \end{center}
\vspace{-0.5cm}
\caption{Braneworld-inspired black holes of Eq.~(\ref{e-rs}).
Allowed regions in the parameter space $(a_*,\beta/M^2)$ for the BH 
candidates GRS 1915+105 and XTE J1550-564 (left panel) and 
GRO J1655-40 and A0620-00 (right panel). The red solid curve 
separates BHs from naked singularities. See text for details.}
\label{f-eta}
\end{figure*}

\begin{figure*}
\begin{center}  
\includegraphics[type=pdf,ext=.pdf,read=.pdf,width=8cm]{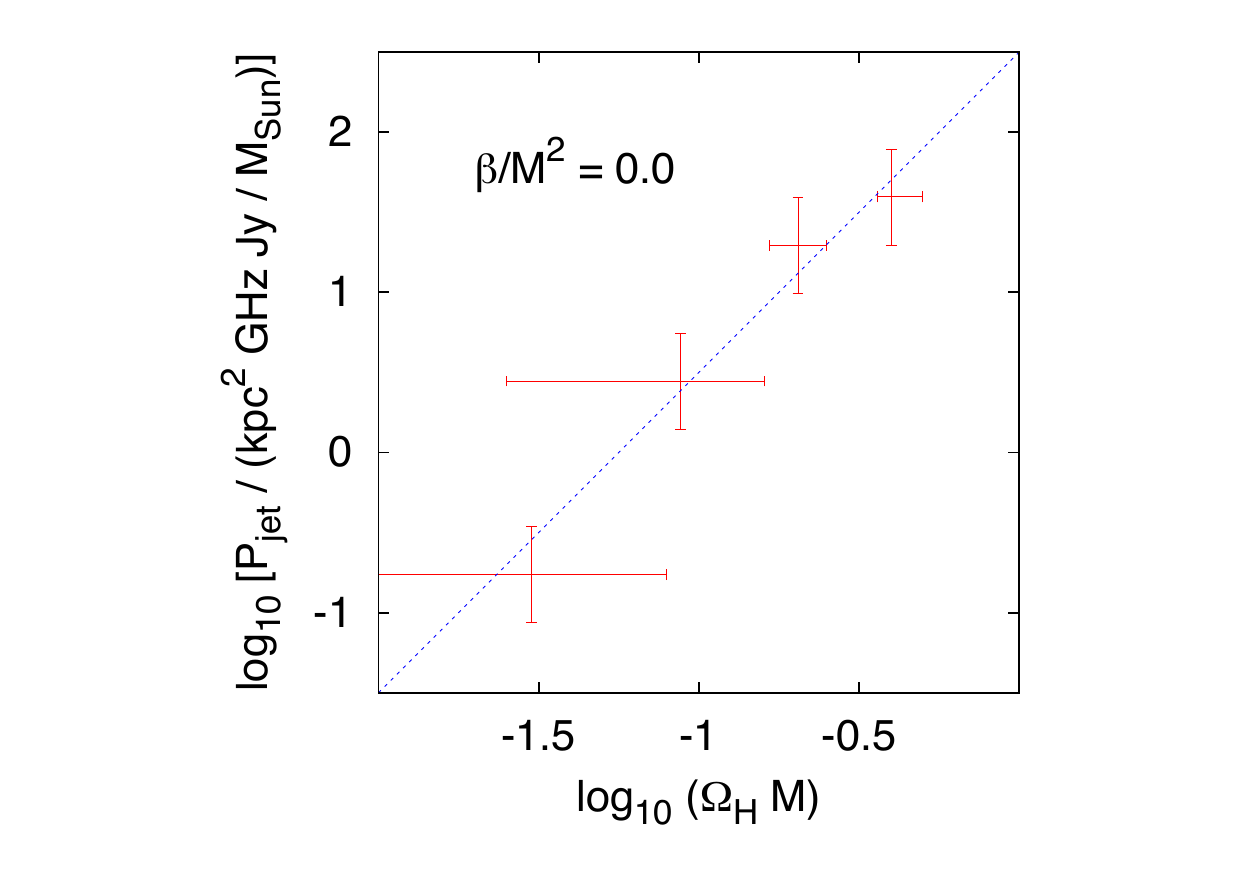}
\includegraphics[type=pdf,ext=.pdf,read=.pdf,width=8cm]{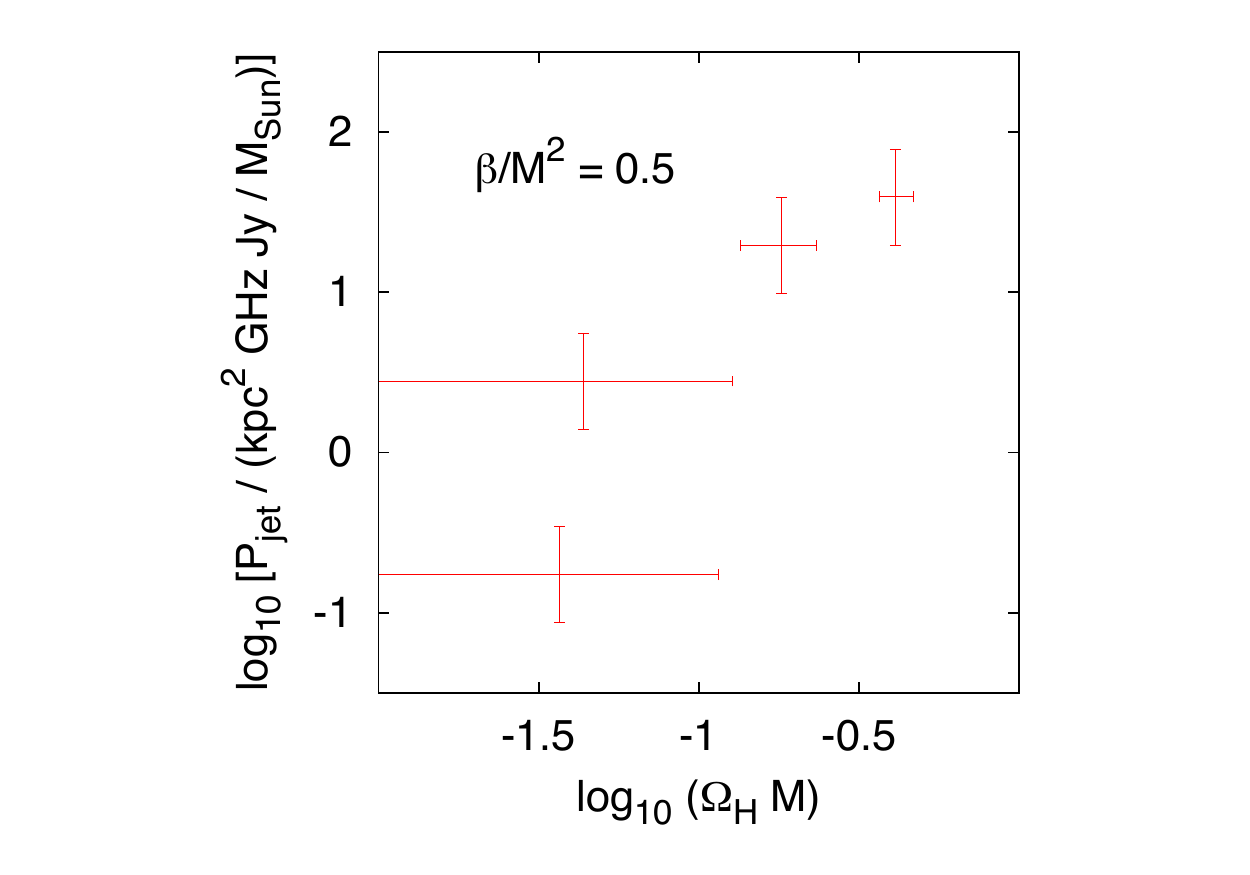} \\
\includegraphics[type=pdf,ext=.pdf,read=.pdf,width=8cm]{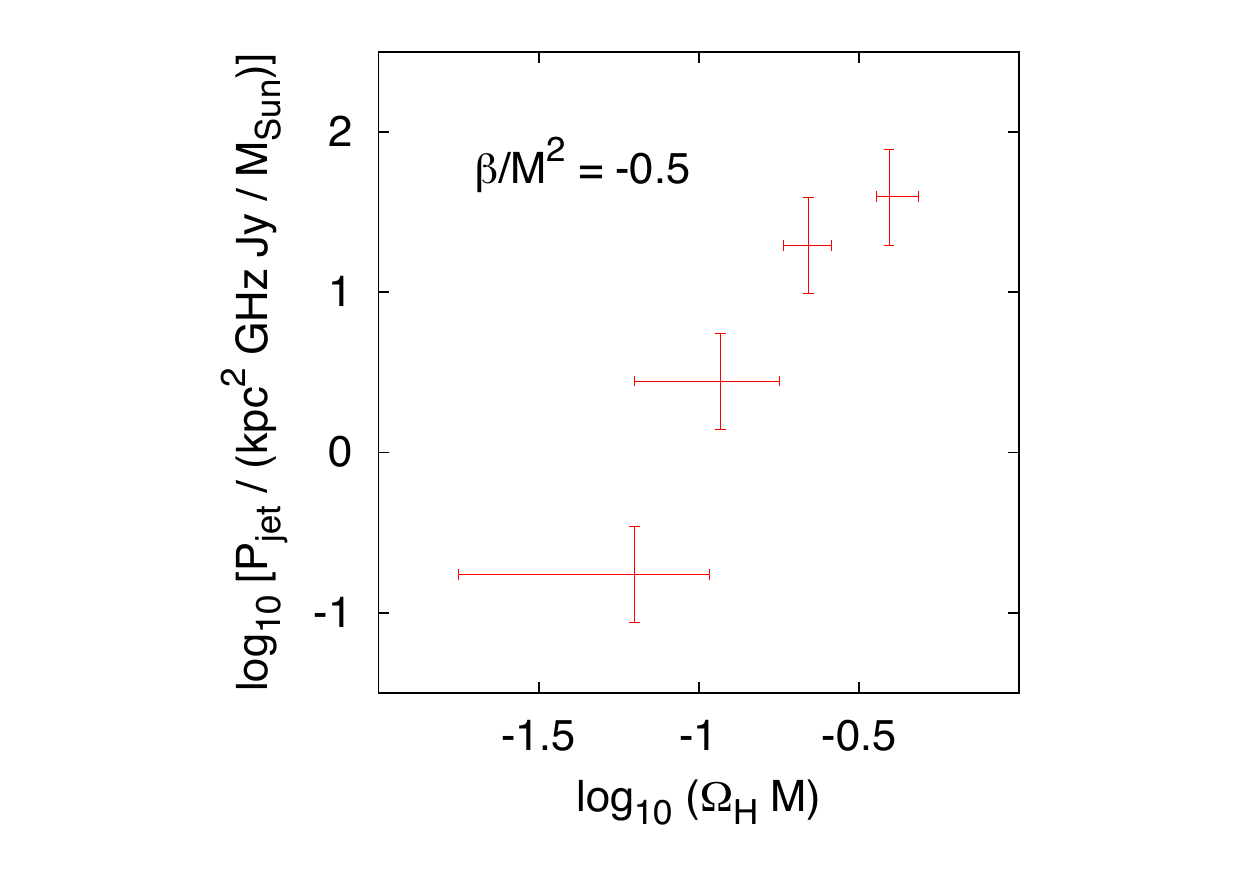}
\includegraphics[type=pdf,ext=.pdf,read=.pdf,width=8cm]{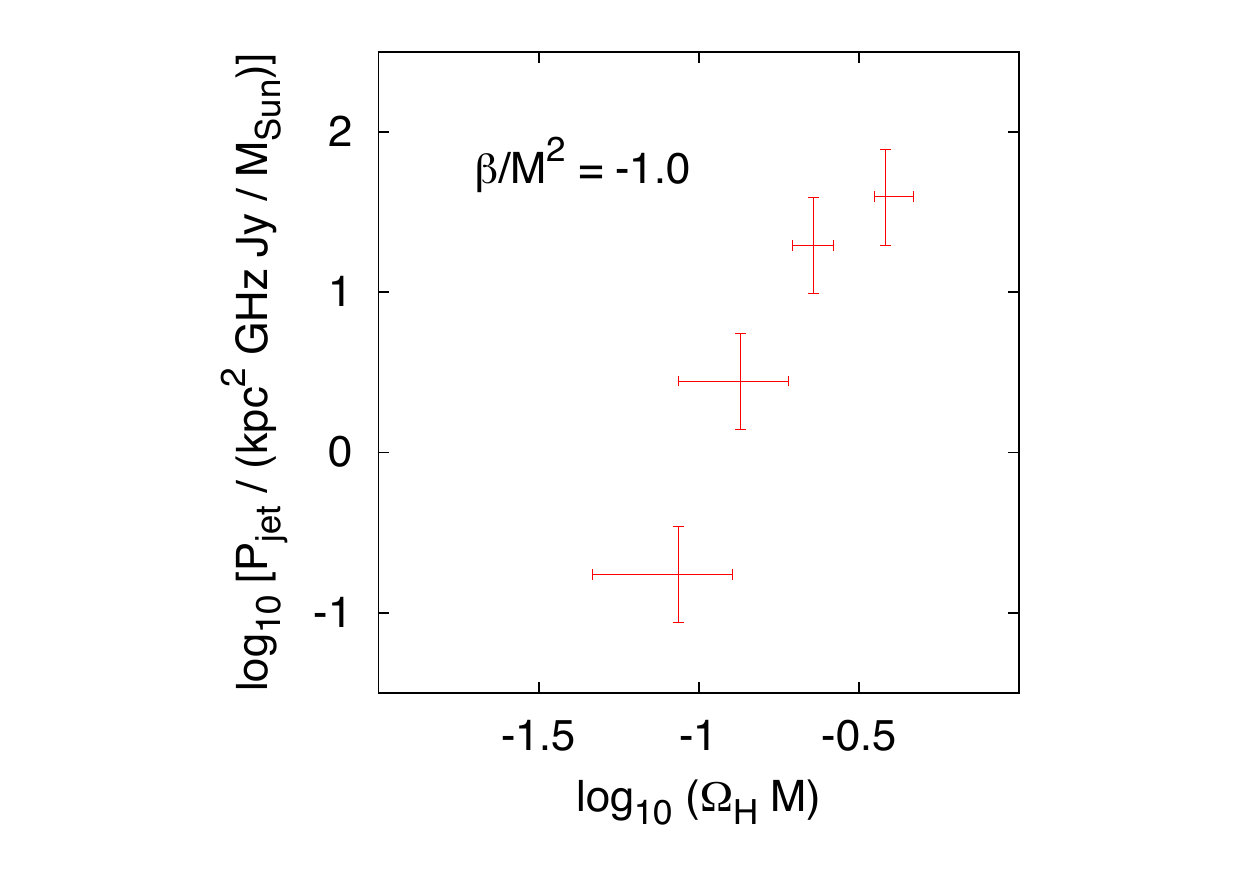} \\
\includegraphics[type=pdf,ext=.pdf,read=.pdf,width=8cm]{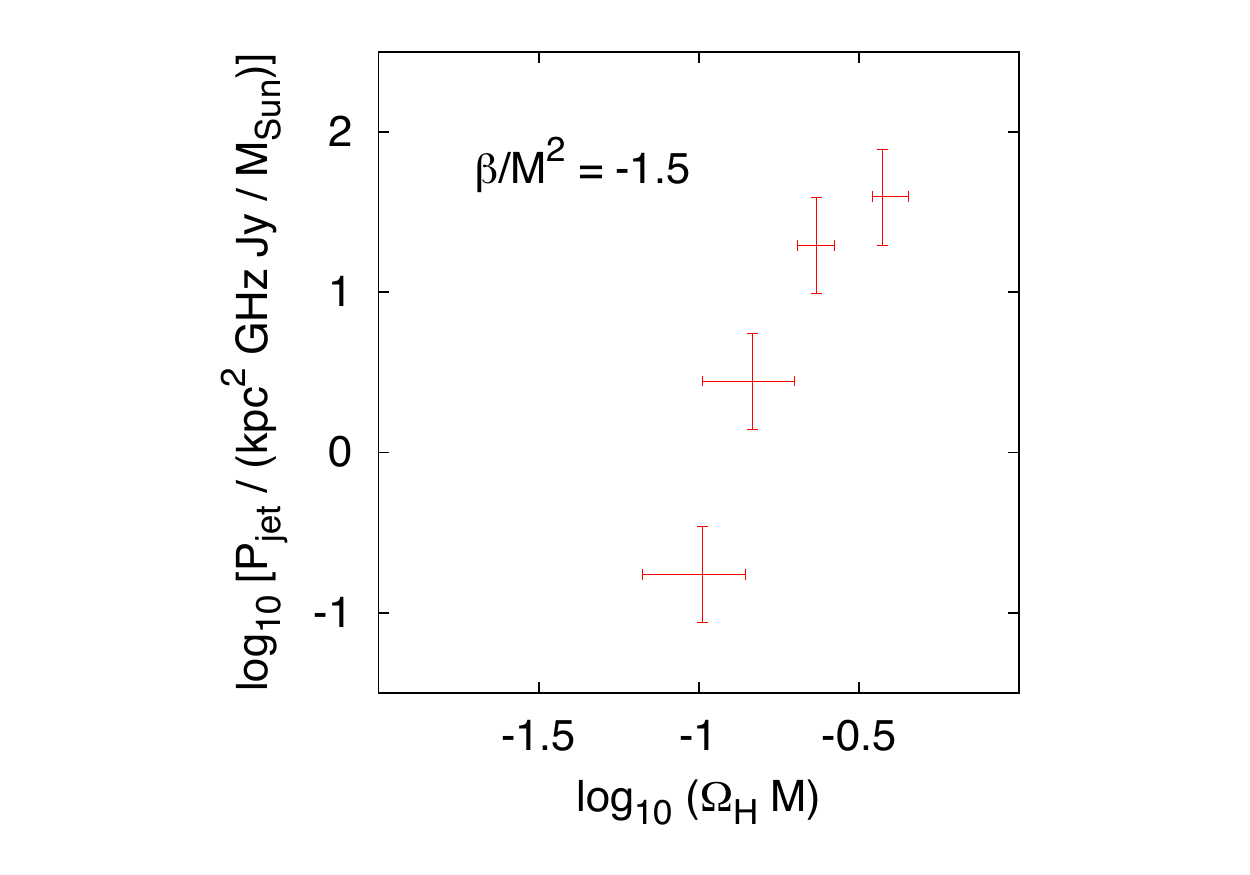}
\includegraphics[type=pdf,ext=.pdf,read=.pdf,width=8cm]{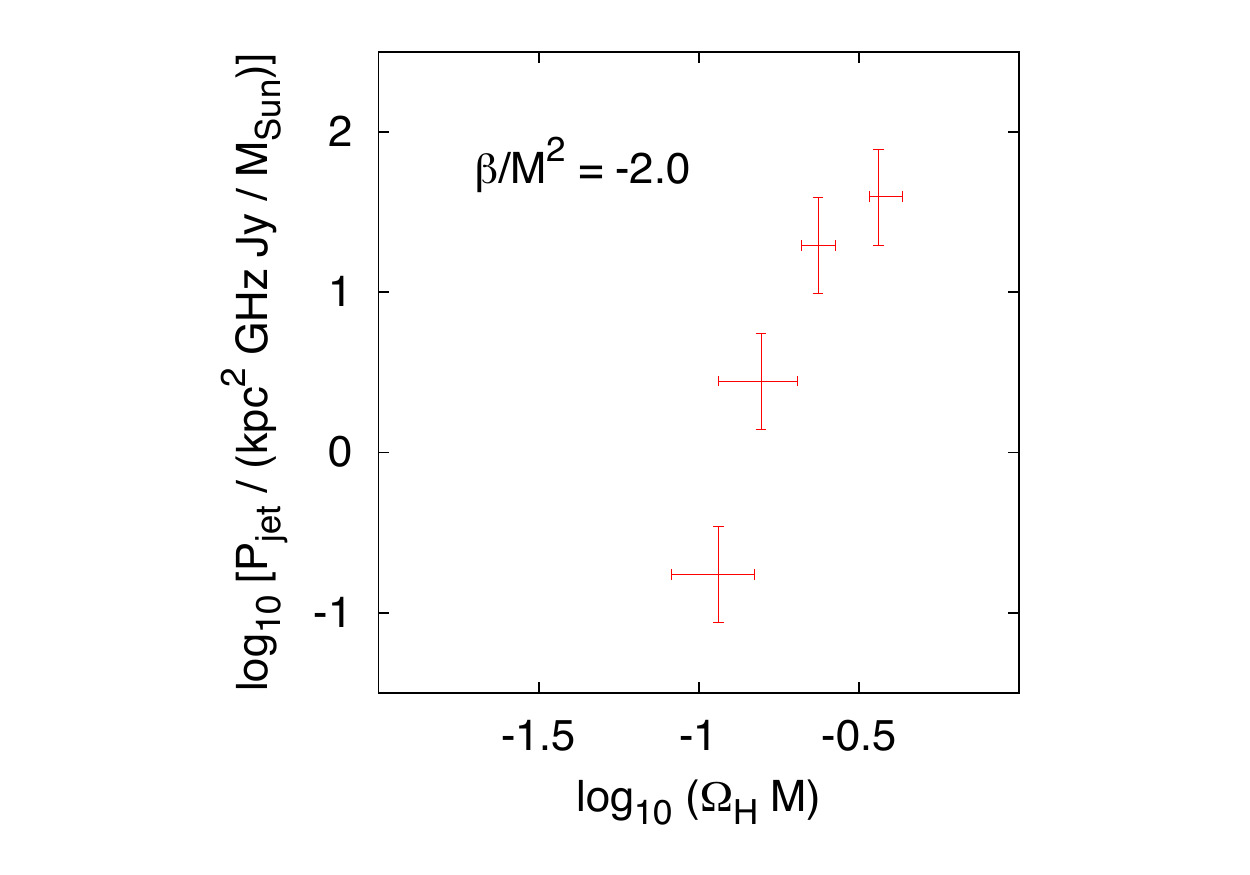}
 \end{center}
 \vspace{-0.5cm}
\caption{Braneworld-inspired black holes of Eq.~(\ref{e-rs}). Plots of 
the jet power $P_{\rm jet}$ against the BH angular frequency $\Omega_H$. 
The top left panel shows the data in the case of the familiar Kerr 
background and the blue dotted line corresponds to $P_{\rm jet} \sim 
\Omega_H^2$, the theoretical scaling derived in Ref.~\cite{tnm}.}
\label{f-jet}
\end{figure*}

\begin{figure*}
\begin{center}  
\includegraphics[type=pdf,ext=.pdf,read=.pdf,width=8cm]{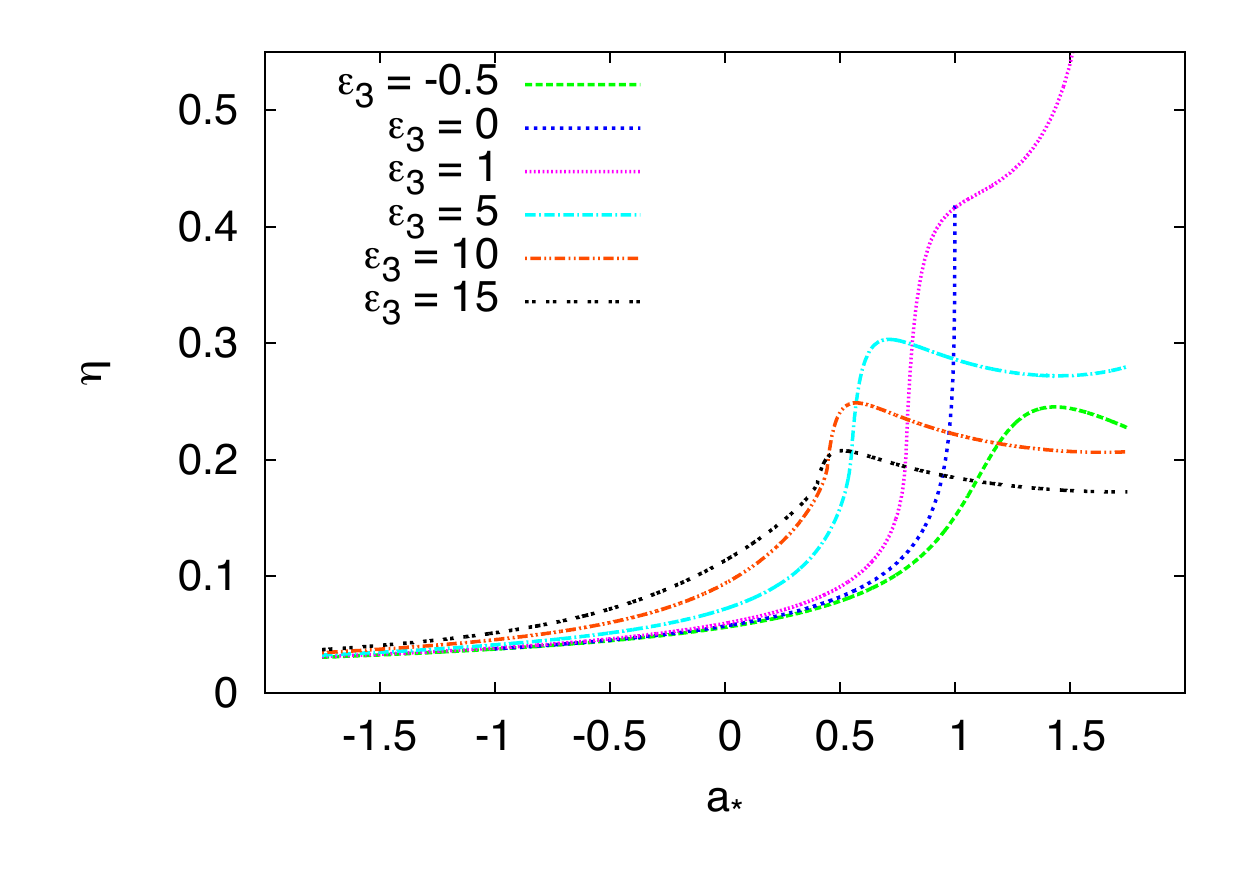}
\vspace{-0.5cm}
\caption{JP black holes of Eq.~(\ref{e-jp}) with deformation parameter
$\epsilon_3$ and $\epsilon_i = 0$ for $i \neq 3$.
Accretion efficiency $\eta = 1 - E_{\rm ISCO}$ as a function 
of the spin parameter $a_*$ for different values of $\epsilon_3$.}
\label{f-om2}
\includegraphics[type=pdf,ext=.pdf,read=.pdf,width=8cm]{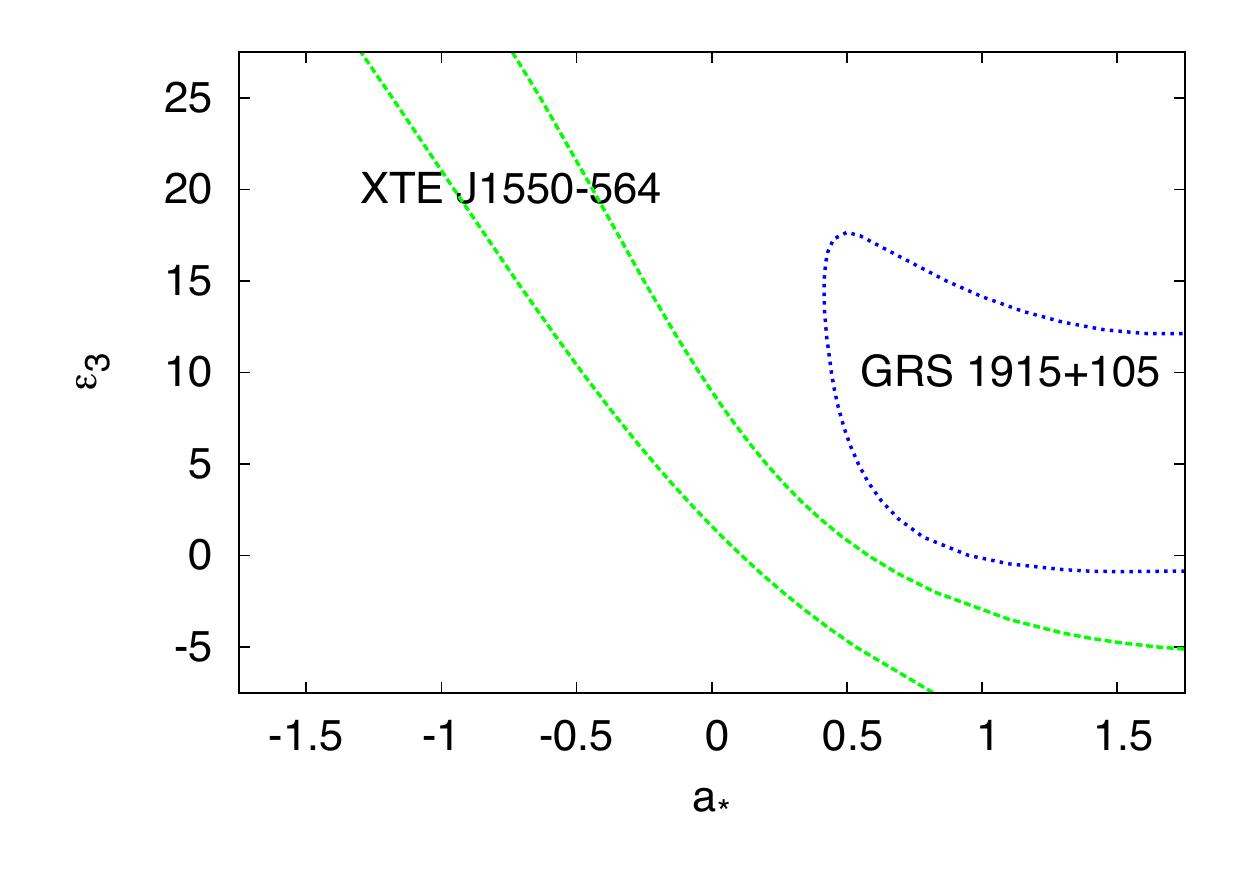}
\includegraphics[type=pdf,ext=.pdf,read=.pdf,width=8cm]{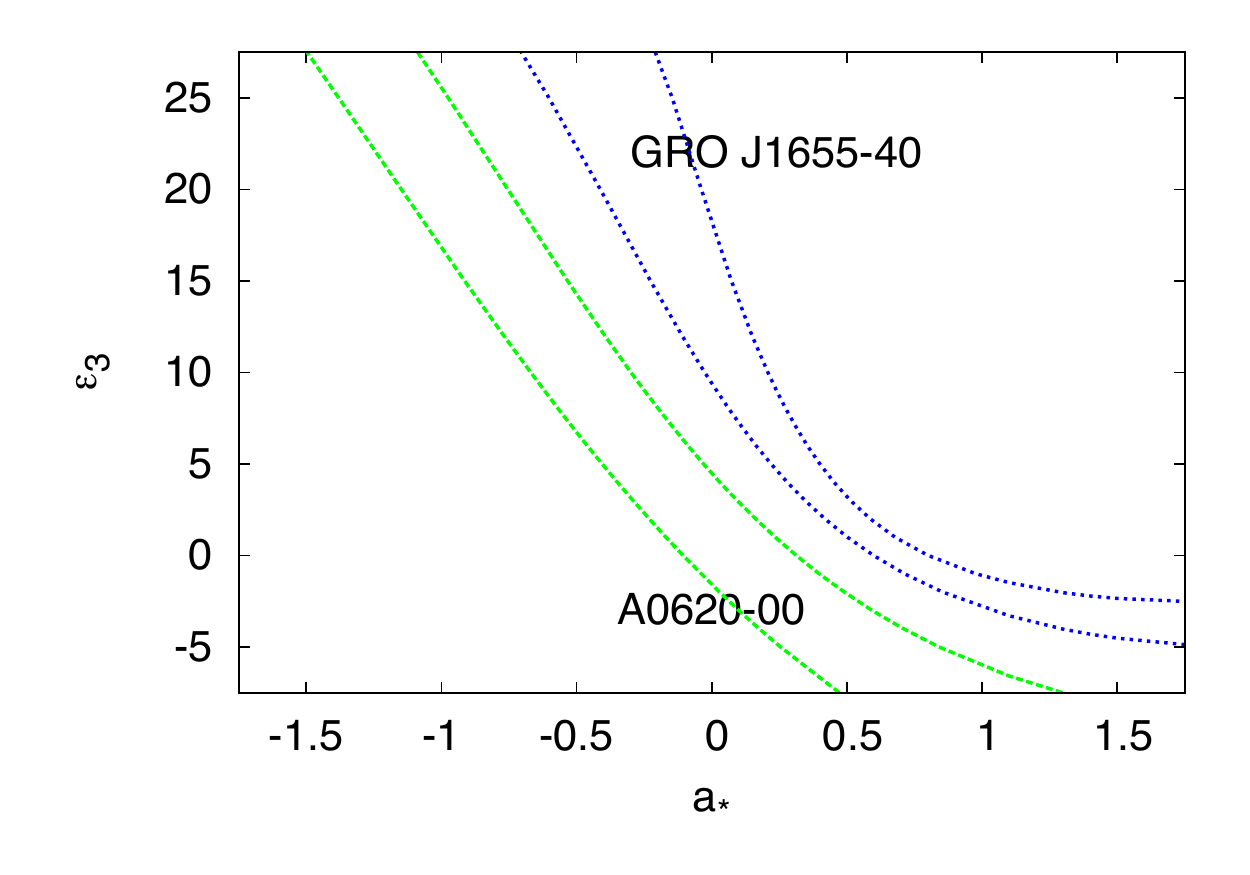}
 \end{center}
\vspace{-0.5cm}
\caption{JP black holes of Eq.~(\ref{e-jp}) with deformation parameter
$\epsilon_3$ and $\epsilon_i = 0$ for $i \neq 3$.
Allowed regions in the parameter space $(a_*,\epsilon_3)$ for the BH 
candidates GRS 1915+105 and XTE J1550-564 (left panel) and 
GRO J1655-40 and A0620-00 (right panel). See text for details.}
\label{f-eta2}
\end{figure*}

\begin{figure*}
\begin{center}  
\includegraphics[type=pdf,ext=.pdf,read=.pdf,width=8cm]{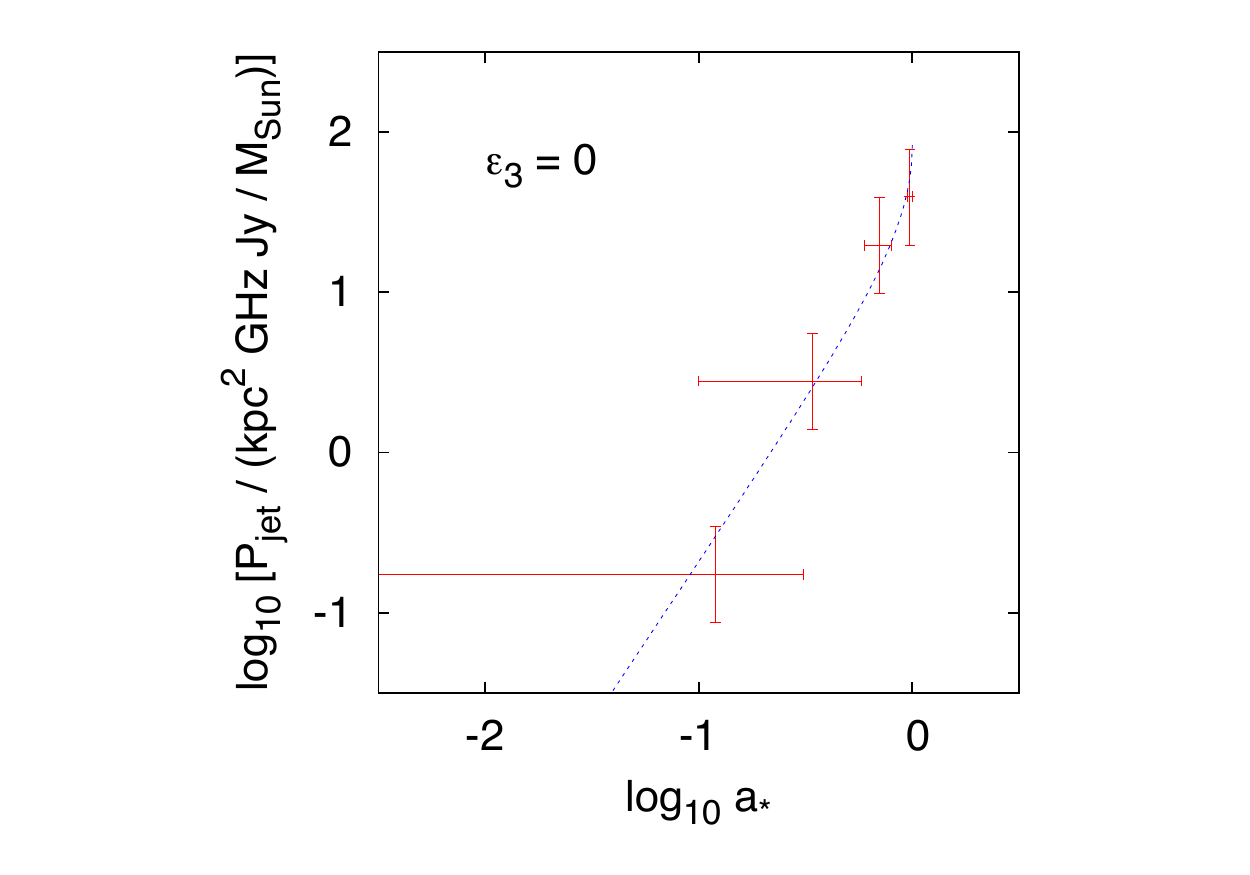}
\includegraphics[type=pdf,ext=.pdf,read=.pdf,width=8cm]{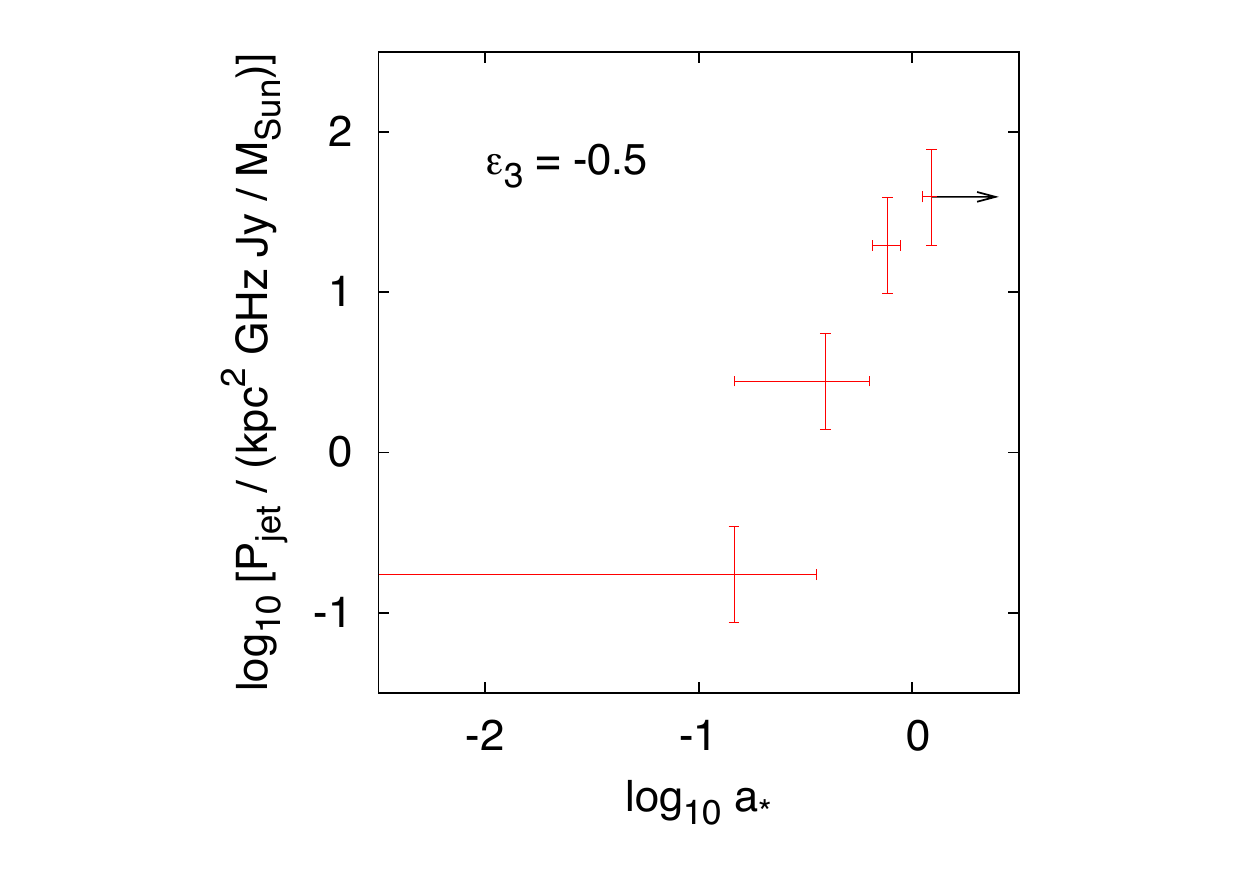} \\
\includegraphics[type=pdf,ext=.pdf,read=.pdf,width=8cm]{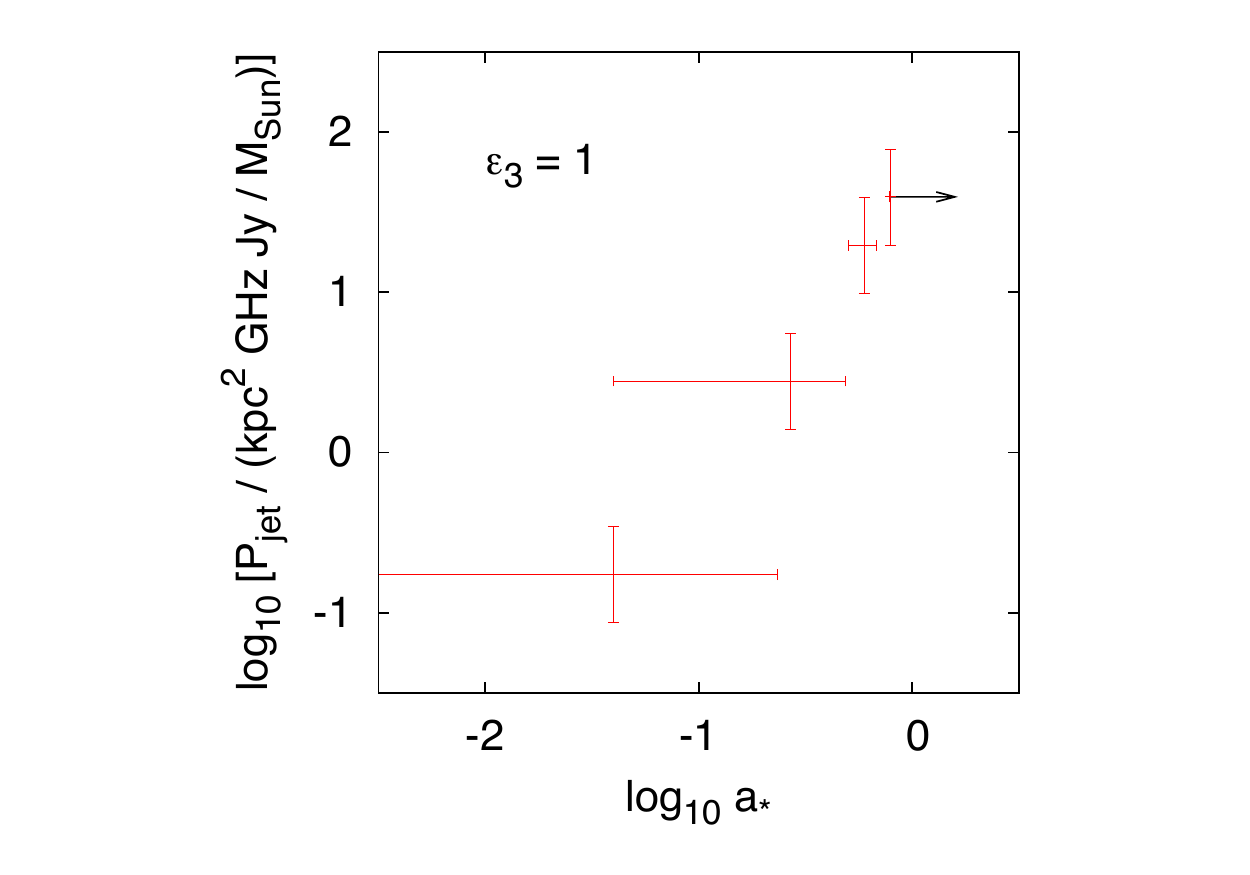}
\includegraphics[type=pdf,ext=.pdf,read=.pdf,width=8cm]{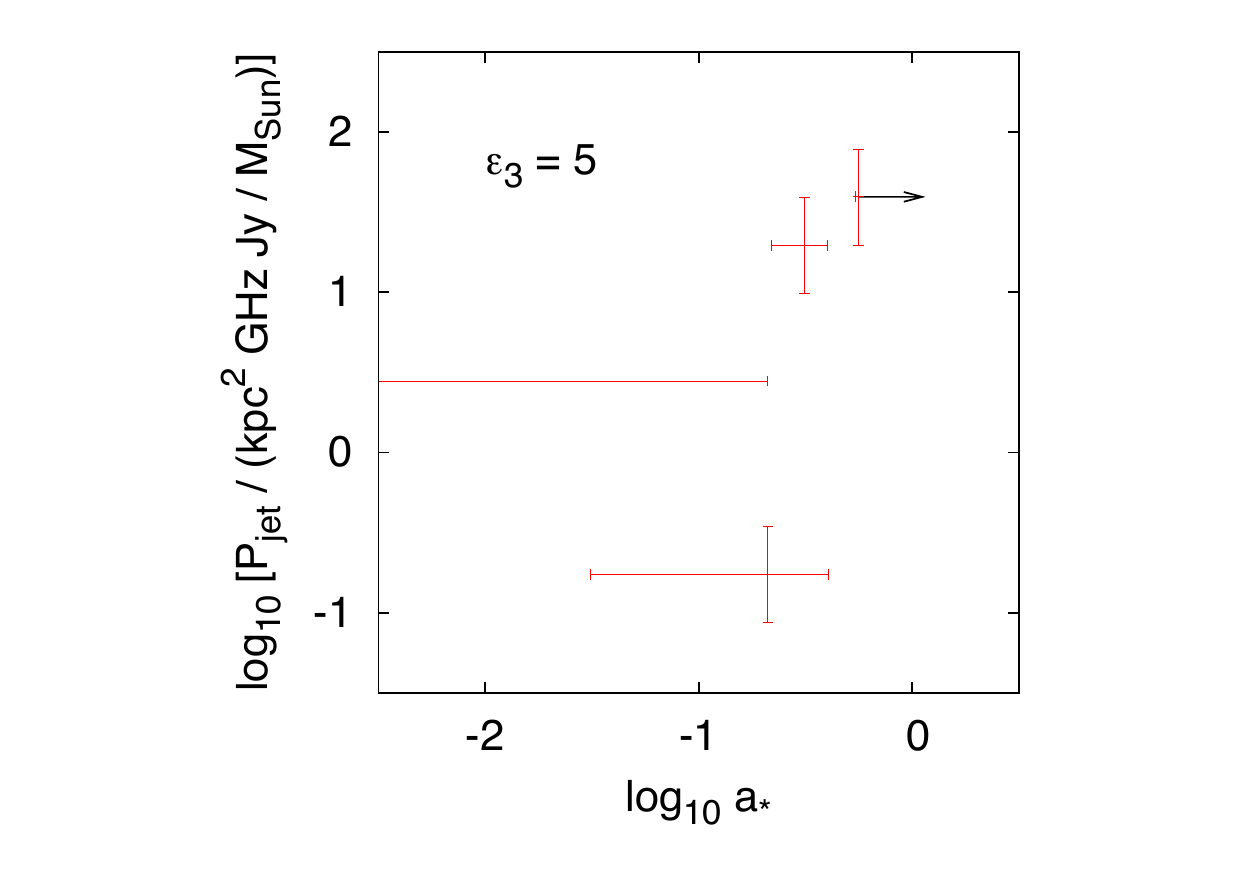} \\
\includegraphics[type=pdf,ext=.pdf,read=.pdf,width=8cm]{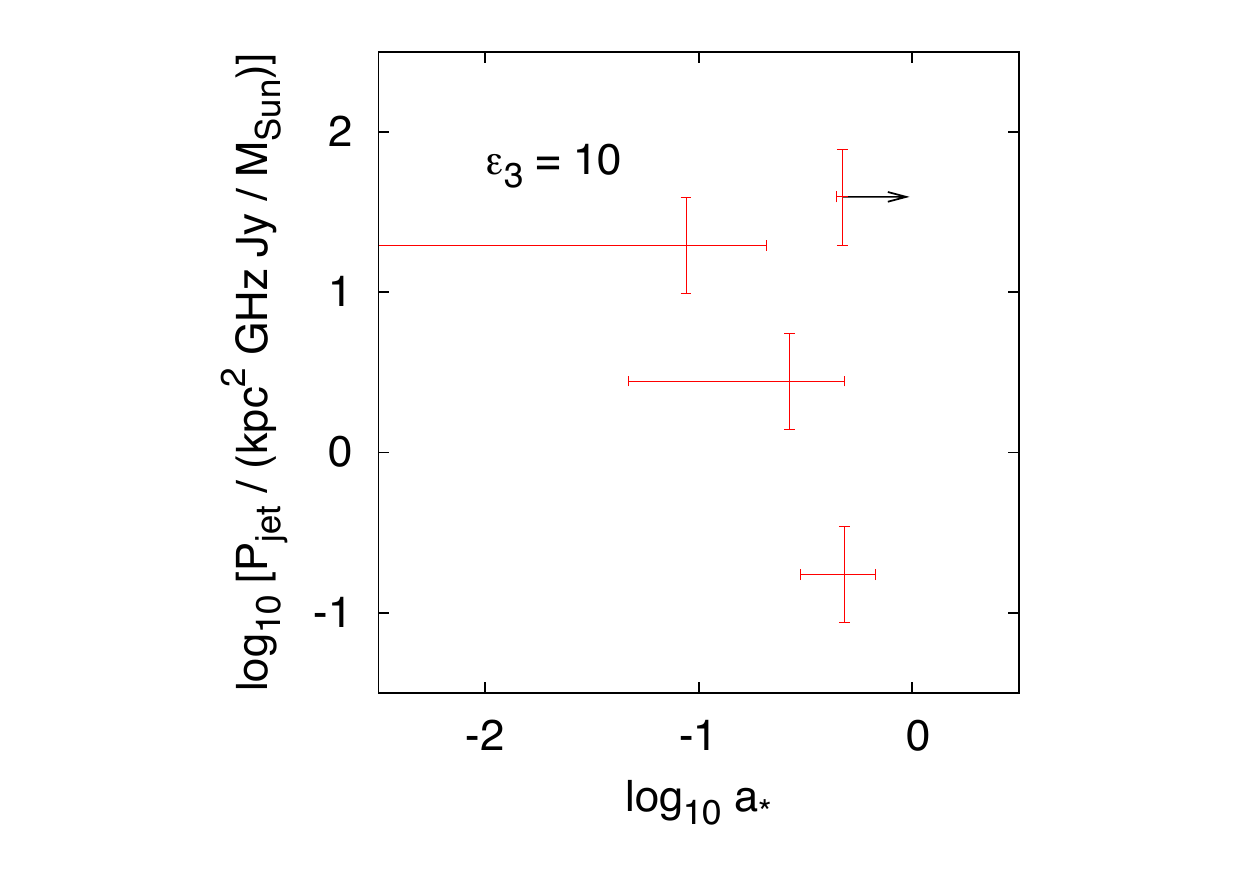}
\includegraphics[type=pdf,ext=.pdf,read=.pdf,width=8cm]{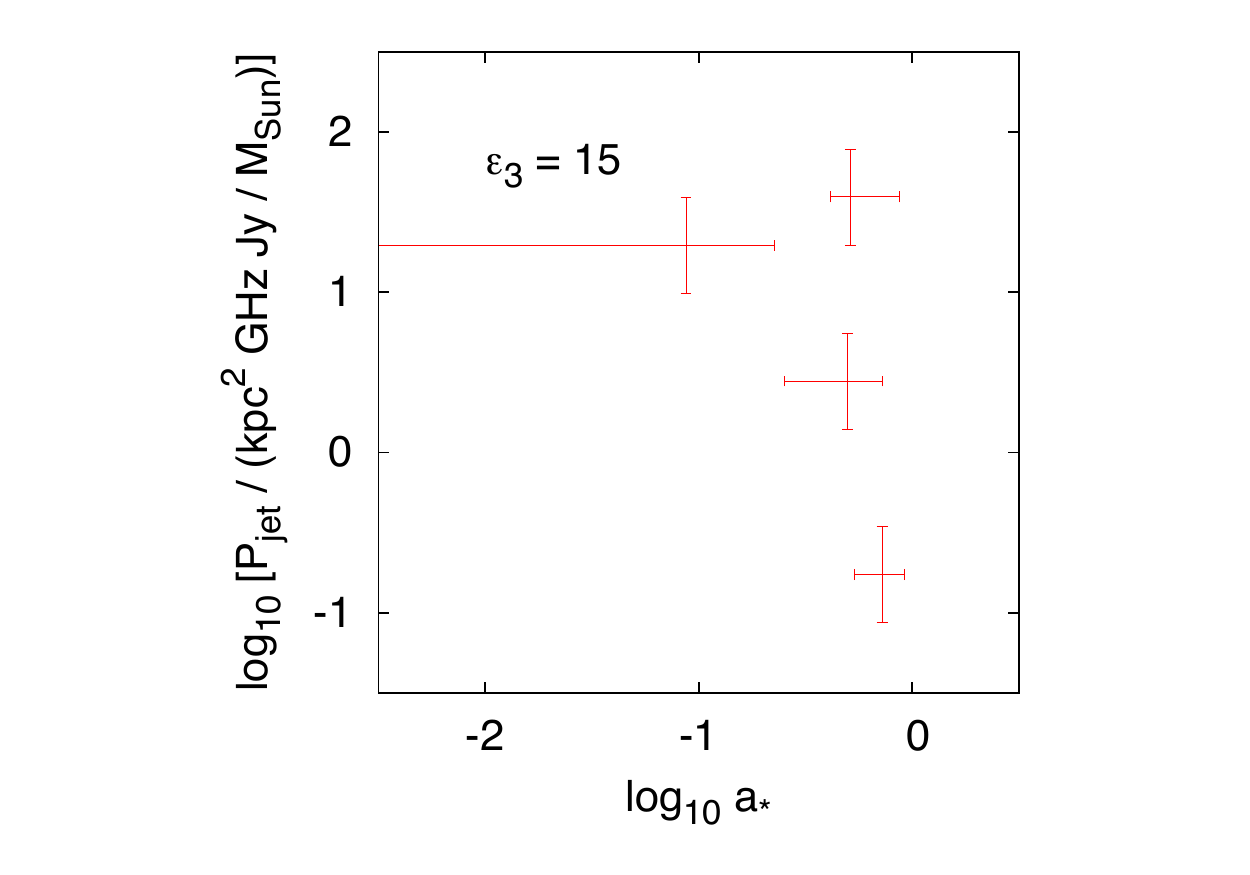}
 \end{center}
 \vspace{-0.5cm}
\caption{JP black holes of Eq.~(\ref{e-jp}) with deformation parameter
$\epsilon_3$ and $\epsilon_i = 0$ for $i \neq 3$. Plots of the jet power 
$P_{\rm jet}$ against the BH spin parameter $a_*$. The top left panel 
shows the data in the case of the familiar Kerr background and the 
blue dotted curve corresponds to $P_{\rm jet} \sim \Omega_H^2$, 
the theoretical scaling derived in Ref.~\cite{tnm}.}
\label{f-jet2}
\end{figure*}

\section{Non-Kerr space-times}

I this section, I will show that the jet power of a BH candidate can provide
additional information about the nature of the compact object and 
potentially can be used to break the degeneracy between the spin and the 
deformation parameter. I will outline the basic idea, without following a 
rigorous study: the latter would require a complete reanalysis of the 
X-ray continuum spectrum of the four objects and new GRMHD simulations 
in a particular non-Kerr background, both beyond the purpose of this 
work, as well as more observational data, which we do not have yet. I will
consider two specific non-Kerr space-times: the braneworld-inspired
BHs of Ref.~\cite{ag} and the Johannsen-Psaltis (JP) BHs of Ref.~\cite{jp}.
These space-times can be seen as the two prototypes of non-Kerr
background, or at least of the ones proposed in the literature~\cite{other}.

\subsection{Example 1: braneworld black holes}

A braneworld-inspired BH solution was found in Ref.~\cite{ag}. In 
Boyer-Lindquist coordinates, the non-zero components of the induced 
4D metric are
\be\label{e-rs}
g_{tt} &=& - \left(1 - \frac{2 M r - \beta}{\rho^2}\right) \, , \nonumber\\
g_{t\phi} &=& - \frac{2 a (2 M r - \beta)}{\rho^2} \sin^2\theta \, , \nonumber\\
g_{\phi\phi} &=& \left[r^2 + a^2 + \frac{2 M r - \beta}{\rho^2} 
a^2 \sin^2\theta \right] \sin^2\theta \, , \nonumber\\
g_{rr} &=& \frac{\rho^2}{\Delta} \, , \nonumber\\
g_{\theta\theta} &=& \rho^2 \, ,
\ee
where
\be
\rho^2 &=& r^2 + a^2 \cos^2\theta \, , \nonumber\\
\Delta &=& r^2 - 2 M r + a^2 + \beta \, ,
\ee
and $\beta$ is the tidal charge parameter, encoding the imprints of the
non-local effects from the extra dimension. The metric looks like the
usual Kerr-Newman solution of General Relativity, which describes a
rotating BH with electric charge $Q$, with $\beta = Q^2$. However,
here $\beta$ can be either positive or negative. The event horizon is 
defined by $\Delta = 0$; the radius of the outer event horizon is
\be
r_H = M + \sqrt{M^2 - a^2 - \beta} \, .
\ee
The event horizon exists only for $M \ge \sqrt{a^2 - \beta}$. When
$M < \sqrt{a^2 - \beta}$, there is no horizon and the space-time has a naked 
singularity\footnote{Let us notice that these braneworld BHs may violate the
familiar bound $|a_*| \le 1$, without violating the weak cosmic censorship
conjecture. It is also possible to check that there exist astrophysical 
processes capable of producing such fast-rotating objects~\cite{a1}.}. 
For the metric in Eq.~(\ref{e-rs}) it is straightforward to repeat the 
analytical derivation of the jet power (see Ref.~\cite{mg} and 
Appendix~A of~\cite{tnm}) and one still finds $P_{\rm jet} \sim \Omega_H^2$, 
as in Kerr.

The analysis of the X-ray continuum spectra of the four objects in
Tab.~\ref{tab} would provide a constraint on $a_*$ and 
$\beta/M^2$\footnote{If the Birkhoff's Theorem holds, Solar System 
experiments would require $|\beta/M^2| < 4.6 \cdot 10^{-4}$. While it 
is not clear if this is the case in braneworld models, the aim of this 
paper is not to constrain these theories, but to show how two 
independent measurements can break the degeneracy between 
the spin and the deformation parameter.}. The
correct procedure would be to reanalyze the X-ray data of these 
objects in the background~(\ref{e-rs}); however, that would take a 
long time and is beyond the purpose of the present paper.
A simple estimate can be obtained from the following consideration.
In the standard case of Kerr background, the continuum-fitting method
provides the BH spin parameter $a_*$ and its mass accretion rate $\dot{M}$,
when the BH mass, its distance from us, and the inclination angle of the
disk are known. Actually, the low frequency region of the spectrum
constrains $\dot{M}$~\cite{dl}, while the position of the peak constrains
the accretion efficiency $\eta = 1 - E_{\rm ISCO}$~\cite{bb}, where 
$E_{\rm ISCO}$ is the specific energy of the gas at the innermost stable 
circular orbit (ISCO), which is supposed to be the inner edge of the 
accretion disk. The common statement in the literature that the 
continuum-fitting method measures the inner radius of the disk, $r_{\rm in}$, 
is correct because in the Kerr metric there is a one-to-one correspondence
between $\eta$ and $r_{\rm in}$. However, in a non-Kerr background 
one can see that the actual key-parameter is $\eta$. We can then write the present
estimates of $a_*$ of the four objects in terms of the accretion efficiency
$\eta$ (see the third column in Tab.~\ref{tab}), and then get the allowed
regions in the space $(a_*,\beta/M^2)$ for every BH candidate (see App.~\ref{app} 
for more details). The accretion efficiency and the BH angular frequency
as a function of the spin parameter are shown in Fig.~\ref{f-om}. The final
results are reported in Fig.~\ref{f-eta}, where the red solid curve separates BHs
from naked singularities. The region with naked singularities can be
excluded for at least two reasons: these space-times have equatorial
stable circular orbits with negative energy, which would imply $\eta > 1$,
and they are presumably unstable, due to the ergoregion instability~\cite{e}.
As we can see in Fig.~\ref{f-eta}, we cannot estimate $a_*$ and $\beta/M^2$
independently, but we can only constrain a combination of the these two
parameters. This is the usual situation we find when we want to test
the Kerr BH hypothesis.

For braneworld BHs, $\Omega_H$ is still given by Eq.~(\ref{e-o}). 
It is also important to notice that $P_{\rm jet}$ is proportional to the second 
power of $\Omega_H$; that is, $P_{\rm jet}$ does not depend on the sense 
of BH rotation with respect to the one of the disk. In Fig.~\ref{f-jet}, I plot the
power jet against $\Omega_H$ for some values of $\beta/M^2$. Here
I assume that all the BH candidates have the same value of $\beta/M^2$.
This assumption can be relaxed and tested when more data will be available.

\subsection{Example 2: JP black holes}

The JP BHs have been proposed in~\cite{jp} explicitly to be used to
test the Kerr BH hypothesis. The non-vanishing metric coefficients in 
Boyer-Lindquist coordinates are:
\be\label{e-jp}
g_{tt} &=& - \left(1 - \frac{2 M r}{\rho^2}\right) (1 + h)
 \, , \nonumber\\
g_{t\phi} &=& - \frac{2 a M r \sin^2\theta}{\rho^2} 
(1 + h) \, , \nonumber\\
g_{\phi\phi} &=& \sin^2\theta \left[r^2 + a^2
+ \frac{2 a^2 M r \sin^2\theta}{\rho^2} \right] + \nonumber\\
&& + \frac{a^2 (\rho^2 + 2 M r) \sin^4\theta}{\rho^2} 
h \, , \nonumber\\
g_{rr} &=& \frac{\rho^2 (1 + h)}{\Delta + 
a^2 h \sin^2\theta } \, , \nonumber\\
g_{\theta\theta} &=& \rho^2 \, ,
\ee
where
\be
\rho^2 &=& r^2 + a^2 \cos^2\theta \, , \nonumber\\
\Delta &=& r^2 - 2 M r + a^2 \, , \nonumber\\
h &=& \sum_{k = 0}^{\infty} \left(\epsilon_{2k} 
+ \frac{M r}{\rho^2} \epsilon_{2k+1} \right)
\left(\frac{M^2}{\rho^2}\right)^k \, .
\ee
The metric has an infinite number of free parameters $\epsilon_i$ and 
the Kerr solution is recovered when all these parameters are set to zero. 
However, in order to recover the correct Newtonian limit we have to 
impose $\epsilon_0 = \epsilon_1 = 0$, while $\epsilon_2$ is constrained 
at the level of $10^{-4}$ from current tests in the Solar System~\cite{jp}.
For the sake of simplicity, in what follows I will consider only the case
with the deformation parameter $\epsilon_3$ and $\epsilon_i = 0$ for 
$i \neq 3$.

For some values of the deformation parameters, the JP BHs have a few
properties common to other non-Kerr metrics, but absent in the Kerr
solution (existence of vertically unstable circular orbits on the equatorial 
plane, topologically non-trivial event horizons, etc.). In particular, here
we cannot define the BH angular frequency, at least in the usual way, as
from Eq.~(\ref{e-o}) we would obtain something that depends on the
polar angle $\theta$. Anyway, if we want to check the Kerr-nature of
astrophysical BH candidates, we can still plot $P_{\rm jet}$ against the
spin parameter $a_*$ and see if the correlation if the one expected
for Kerr BHs.

The accretion efficiency $\eta = 1 - E_{\rm ISCO}$ as a function of the
spin parameter $a_*$ for some values of the deformation parameter
$\epsilon_3$ is shown in Fig.~\ref{f-om2}. To get the constraints on
$a_*$ and $\epsilon_3$ for the four objects in Tab.~\ref{tab}, we can 
still apply the simplified analysis of the previous subsection. The
results are shown in Fig.~\ref{f-eta2}. Fig.~\ref{f-jet2} shows the plots
$P_{\rm jet}$ vs $a_*$ in the JP space-time with $\epsilon_3$. The
blue-dotted curve in the top left panel is the theoretical scaling
$P_{\rm jet} \sim \Omega_H^2$ in Kerr background. Let us notice that
the cases with $\epsilon_3 = 10$ and 15 are allowed with the sole use
of the continuum-fitting method, while they seem to be at least strongly
disfavored when we add the information coming from the jet power. Indeed,
when $\epsilon_3 = 10$ and 15, the continuum-fitting method would 
predict a counterrotating disk (i.e. $a_* < 0$) for  some sources, while the 
jet power should be independent of the sense of BH rotation
with respect to the accreting matter.

\section{Conclusions}

Astrophysical BH candidates are thought to be the Kerr BHs predicted 
in General Relativity, but direct observational evidence for this identification 
is still lacking. In order to test and verify the Kerr BH hypothesis, we have 
to probe the geometry of the space-time around these objects. The current 
most robust approach to do that with already available data seems to be 
the continuum-fitting method, a technique used by astronomers to measure 
the spin of the stellar-mass BH candidates. The physics involved is relatively 
simple and there are both astrophysical observations and numerical 
calculations supporting the crucial ingredients of this approach. However, 
the continuum-fitting method cannot provide at the same time an estimate 
of the spin and of some deformation parameter measuring the deviations 
from the Kerr geometry. The problem is that there is a degeneracy between 
these two parameters and therefore it is only possible to get a constraint 
on some combination of them. The reason is that the continuum-fitting 
method is sensitive to the accretion efficiency, which depends on the 
spin and on the deformation parameter.

In this paper, I explored a way to break this degeneracy and get an 
estimate of the spin and on the deformation parameter separately. If 
transient ballistic jets in BH binaries are powered by the BH spin via 
the Blandford-Znajek mechanism, the jet power and the BH spin should 
be correlated in a specific way. In Ref.~\cite{nm}, the authors showed 
for the first time evidence for such a correlation. Here, I showed that, if this interpretation 
is correct, the estimate of jet power provides an additional information 
about the nature of the stellar-mass BH candidates and, when combined 
with the continuum-fitting method, it can potentially be used to constrain 
the deformation parameter. As it is particularly clear in Fig.~\ref{f-jet2}, where 
$\epsilon_3$ is the deformation parameter and $\epsilon_3 = 0$ corresponds 
to the Kerr metric, the expected correlation (the blue dotted curve in the top 
left panel of Fig.~\ref{f-jet2}) is not consistent with observations when
the space-time has large deviations from the Kerr solution 
(the cases $\epsilon_3 = 10$ and 15 in Fig.~\ref{f-jet2}). The interpretation
of the authors of Ref.~\cite{nm} needs to be confirmed and the study of
a larger number of objects is compulsory. However, as shown in this
work through a simplified analysis, the combination of the continuum-fitting
method and the estimate of jet power may be able to test the Kerr-nature
of stellar-mass BH candidates in the near future.


\begin{acknowledgments}
This work was supported by the Humboldt Foundation.
\end{acknowledgments}


\appendix

\section{Accretion efficiency in the Novikov-Thorne model \label{app}}

The Novikov-Thorne model is the standard model for accretion 
disks~\cite{nt}. It describes geometrically thin and optically thick disks 
and it is the relativistic generalization of the Shakura-Sunyaev
model~\cite{ss}. Accretion is possible because viscous magnetic/turbulent 
stresses and radiation transport energy and angular momentum 
outwards. The model assumes that the disk is on the equatorial plane 
and that the disk's gas moves on nearly geodesic circular orbits.
The model can be applied for a generic stationary, axisymmetric, and
asymptotically space-time. Here, the line element can always be written 
as
\be
ds^2 = g_{tt} dt^2 + 2g_{t\phi}dt d\phi + g_{rr}dr^2 
+ g_{\theta\theta} d\theta^2 + g_{\phi\phi}d\phi^2 \, .
\nonumber\\
\ee
Since the metric is independent of the $t$ and $\phi$ coordinates, we 
have the conserved specific energy at infinity, $E$, and the conserved 
axial-component of the specific angular momentum at infinity, $L$. 
From the conservation of the rest-mass, $g_{\mu\nu}u^\mu u^\nu = -1$,
we can write
\be
g_{rr}\dot{r}^2 + g_{\theta\theta}\dot{\theta}^2 = V_{\rm eff}(r,\theta) \, ,
\ee
where the effective potential $V_{\rm eff}$ is given by
\be
V_{\rm eff} = \frac{E^2 g_{\phi\phi} 
+ 2 E L g_{t\phi} + L^2 g_{tt}}{
g_{t\phi}^2 - g_{tt} g_{\phi\phi}} - 1 \, .
\ee
Circular orbits in the equatorial plane are located at the zeros and the 
turning points of the effective potential: $\dot{r} = \dot{\theta} = 0$, which 
implies $V_{\rm eff} = 0$, and $\ddot{r} = \ddot{\theta} = 0$, requiring 
respectively $\partial_r V_{\rm eff} = 0$ and $\partial_\theta V_{\rm eff} = 0$.
From these conditions, one can obtain the angular velocity, $E$, and $L$:
\be
\Omega &=& \frac{- \partial_r g_{t\phi} 
+ \sqrt{\left(\partial_r g_{t\phi}\right)^2 
- \left(\partial_r g_{tt}\right) \left(\partial_r 
g_{\phi\phi}\right)}}{\partial_r g_{\phi\phi}} \, , \\
E &=& - \frac{g_{tt} + g_{t\phi}\Omega}{
\sqrt{-g_{tt} - 2g_{t\phi}\Omega - g_{\phi\phi}\Omega^2}} \, , \\
L &=& \frac{g_{t\phi} + g_{\phi\phi}\Omega}{
\sqrt{-g_{tt} - 2g_{t\phi}\Omega - g_{\phi\phi}\Omega^2}} \, .
\ee
The orbits are stable under small perturbations if $\partial_r^2 V_{\rm eff} \le 0$ 
and $\partial_\theta^2 V_{\rm eff} \le 0$. In Kerr space-time, the second 
condition is always satisfied, so one can deduce the radius of the innermost 
stable circular orbit (ISCO) from $\partial_r^2 V_{\rm eff} = 0$. In general, 
however, that may not be true. For instance, in the JP space-times, the ISCO
radius may be determined by the orbital stability along the vertical direction. 
When we know the ISCO radius, we can compute the corresponding
specific energy $E_{\rm ISCO}$ and then the accretion efficiency:
\be
\eta = 1 - E_{\rm ISCO} \, .
\ee


\end{document}